\documentclass[3p,times,procedia]{elsarticle}
\usepackage{nupha_ecrc}

\volume{00}

\firstpage{1}

\journalname{Nuclear Physics A}

\runauth{Md. Rihan Haque}

\jid{nupha}

\jnltitlelogo{Nuclear Physics A}

\usepackage{amssymb}

\begin{document}

\begin{frontmatter}



 \dochead{} 

\title{Measurements of the chiral magnetic effect in Pb--Pb collisions with ALICE}


\author{Md. Rihan Haque for the ALICE Collaboration}

\address{}

\begin{abstract}
We present the measurement of the charge-dependent 3-particle
azimuthal correlation for unidentified charged particles in Pb--Pb
collisions at $\sqrt{s_{\rm NN}}$ = 5.02 TeV in ALICE. The results are
compared with corresponding results from Pb--Pb collisions at
$\sqrt{s_{\rm NN}}$ = 2.76 TeV. We observe no significant difference
in the charge-sensitive 3-particle correlator ($\gamma_{112}$) between
the two collision energies. Charged-dependent mixed-harmonic
correlator ($\gamma_{132}$) is also presented and compared with the
predictions from a blast-wave model incorporating local charge
conservation. 

\end{abstract}


\end{frontmatter}



\section{Introduction}
It has been speculated that a heavy-ion collision can give rise to a
very strong magnetic field $\vec{B}$ ($\sim$m$^{2}_{\pi}$
GeV$^{2}$/$c^{4}$) due to the relativistic motion of charged
nuclei~\cite{MagField}. This magnetic field can interact with the non-trivial
topological charge present in the hot and dense medium of quarks and
gluons (known as QGP), and create a non-zero current along $\vec{B}$.
The end result of this interaction is a charge asymmetry along the
direction of $\vec{B}$. This phenomenon is called the Chiral Magnetic
Effect or CME~\cite{CME_Intro}. 
 On average, $\vec{B}$ is perpendicular to the reaction plane angle,
 $\Psi_{RP}$, which is defined as the angle between the impact
 parameter and the laboratory $x$-axis. Since we cannot measure
 the impact parameter, we use the symmetry plane angle
 ($\Psi_{\rm n}$) for each harmonic n. On average, the 2$^{nd}$ order
 symmetry plane ($\Psi_{2}$) measures the direction of $\Psi_{RP}$. 
 Therefore, any charge difference along $\vec{B}$ can be measured with
 respect to $\Psi_{2}$. The 3-particle correlator to study the CME
 effect was first suggested in~\cite{CME_Gamma}, it can be
 generalized as, 
\begin{equation}
\gamma_{\rm nmp} = \cos[\rm n\phi_{\alpha} + \rm m\phi_{\beta} - (\rm
n+\rm m)\Psi_{\rm k}],
\end{equation} 
where $\phi_{\alpha}$ and  $\phi_{\beta}$ are the azimuthal angles of
two particles from the same event with either opposite or the same
charge  and $\Psi_{\rm k}$ is the $\rm k^{th}$ order symmetry plane. 
The symmetry plane $\Psi_{\rm k}$ is defined as, 
\begin{equation}
\Psi_{\rm k} = \frac{1}{\rm k}\tan^{-1}\frac{Q_{\rm k,y}}{Q_{\rm k,x}}, \\ 
\end{equation} 
where $Q_{\rm k,y} = \sum_{i=1}^{M}\sin(\rm k\phi_{i})$ and $Q_{\rm k,y} =
\sum_{i=1}^{M} \cos(\rm k\phi_{i})$ are the components of the flow 
vector $\vec{Q_{\rm k}}$ for an event with multiplicity M.
The first harmonic $\gamma_{112}$ is sensitive to charge separation due 
to CME as $\Psi_{2}$, on average, is correlated (perpendicular) to
$\vec{B}$.  The leading backgrounds to this correlator are collective
flow ($v_{2}$), weak-decay pairs (of opposite sign) and local charge
conservation (which can be studied in balance function measurements).  
Previous measurements done at RHIC and at the LHC have shown that the  
first harmonic $\gamma_{112}$ has the same strength in both Au--Au and 
Pb--Pb collisions despite the ten fold difference in the collision 
energy~\cite{CME_ALICE_PRL}. A very recent measurement by ALICE
used the event-shape-engineering technique to classify events
according to the $2^{nd}$ order flow vector and put a constrain over the
strength of the CME signal to be in the range 26\% to 33\% at a 95\%
confidence level~\cite{CME_ESE}. In this analysis we follow an
approach to disentangle the background by measuring mixed and higher
harmonics  such as $\gamma_{224}$, $\gamma_{132}$ and $\gamma_{123}$.  
These mixed and higher harmonic correlators are not
sensitive to CME due to the inclusion of higher order symmetry planes 
($\rm k>2$), which do not have any correlation with
$\vec{B}$. Therefore, mixed and higher harmonics would contain mostly
background, which is also present in $\gamma_{112}$ along with the
CME signal. A theoretical model, $e.g.$, blast-wave model
incorporating local charge conservation~\cite{CME_ESE} could be tuned
to reproduce the background as measured by $\gamma_{132}$,
$\gamma_{123}$ and $\gamma_{224}$. Then,
this model can predict the background contribution to $\gamma_{112}$ 
and enable us to quantify the CME signal measured in data. The study
of $\gamma_{112}$ with identified particles will allow us to
investigate the particle-type dependence of the CME.  

\section{The ALICE detector}
The ALICE detector consists of many sub-detector
systems~\cite{ALICE1, ALICE2}. For the analysis with unidentified
charged particles we have used the TPC and the ITS for the tracking
and the V0 detectors (V0A and V0C) for the symmetry-plane calculation
and the centrality estimation. A total of 34 million minimum-bias Pb--Pb
events at $\sqrt{s_{\rm NN}}$ = 5.02 TeV were analyzed. Events with the
$z$-component of the collision vertex lying between
$\pm$10 cm from the nominal position (center of the ALICE tracking
detectors) and at least two charged particles have 
been used for the analysis. A strict selection criteria on the distance
of closest approach (DCA) is applied for each charged particle with
respect to the event vertex to minimize the secondary particles (such
as those from weak decays, particles originating from beam$-$material or 
beam$-$gas events). Charged particles are selected if they have
at least one tracking point originating in the ITS which also helps to
remove secondaries. The correlation on the number of tracks from ITS and
from TPC has been used to remove pile-up events. To take into account 
the detector inefficiency the charge particles are weighted
accordingly. These weights are evaluated using simulated
tracks from Monte-Carlo event generators convoluted with a detector
response.  Finally, the inverse of the azimuthal distribution of the
charged particles (averaged over all events of a particular class)
are used as weights to account for non-uniform azimuthal acceptance
(such as dead/inactive sectors of TPC). The variation of all these
selection criteria is used to determine the systematic uncertainties
in the analysis.

\section{Results and Discussion}

\begin{figure}[!ht]  
 \centering
  \includegraphics[totalheight=6.0cm]{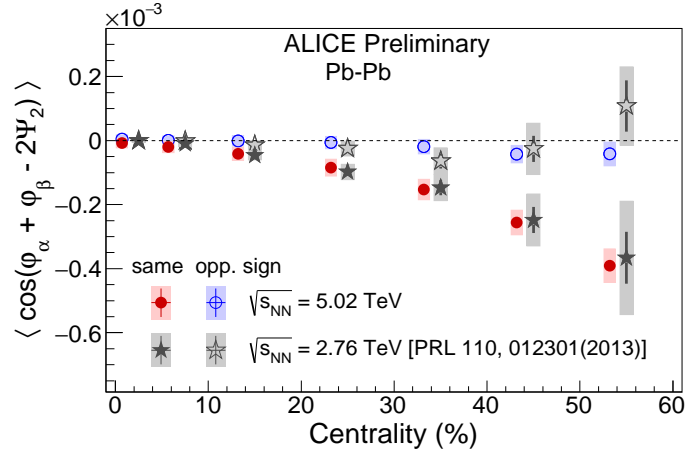}
  \caption{\small{Opposite-sign (red) and same-sign (blue)
      $\gamma_{112}$ vs. centrality. The error bars and
      bands on each marker corresponds to statistical and systematic
      uncertainty, respectively.}}
\label{fig:C112vsCent}
 \end{figure}

\begin{figure}[!ht] 
 \centering
  \includegraphics[totalheight=6.0cm]{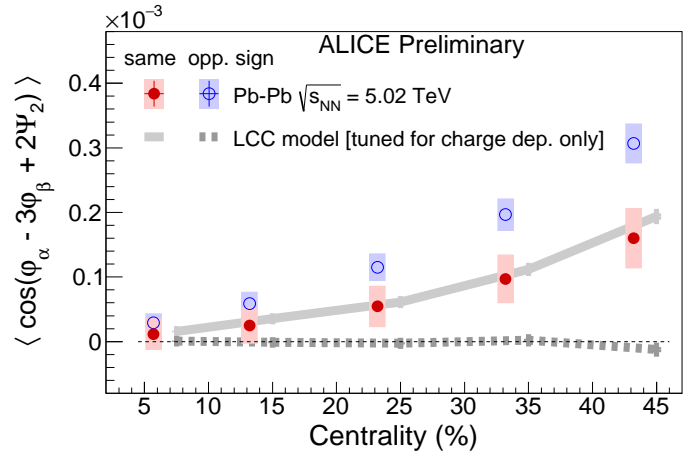}
  \caption{\small{Opposite-sign (red) and same-sign (blue)
      $\gamma_{132}$ vs. centrality. The error bars and
      bands on each marker corresponds to statistical and systematic
      uncertainty, respectively. The solid lines corresponds to the
      model predictions (see text for details).}}
 \label{fig:C132vsCent}
 \end{figure}

Figure~\ref{fig:C112vsCent} shows the $\gamma_{112}$ correlator as a
function of centrality measured in Pb--Pb collisions at $\sqrt{s_{\rm
 NN}}$ = 5.02 TeV. The measurements are also compared with results
from Pb--Pb collisions at $\sqrt{s_{\rm NN}}$ = 2.76 TeV~\cite{CME_ALICE_PRL}.
  Despite the two fold increase in the center-of-mass energy we
do not see any significant change in the magnitude of the correlator
for the same-sign and opposite-sign pairs. We can observe from other
measurements that the background to the $\gamma_{112}$ correlator
(such as elliptic flow $v_{2}$, and width of the balance function)
does not change much for these two colliding
energies~\cite{v2_PbPb2and5,BF_Charge}. 
Therefore, the agreement of $\gamma_{112}$ in these two colliding
energies suggests that there is no apparent enhancement of the CME
signal.  Figure~\ref{fig:C132vsCent} shows the mixed harmonic $\gamma_{132}$
correlator as function of centrality measured in Pb--Pb collisions at
$\sqrt{s_{\rm NN}}$ = 5.02 TeV. The data dave been compared with a
Blast-wave + LCC model and shown as the continuous thick lines. The
model clearly under predicts the opposite sign results,  which is to
be expected as the model does not include all possible background
sources ($e.g.$ the model does not include weak decays, resonances). Further
improvement to the model could give a better agreement to reproduce
the measured mixed and higher harmonic correlators. The development of
the model and comparison of the model predictions with other mixed and
higher harmonics ($e.g.$ $\gamma_{123}$ and $\gamma_{224}$) will be
investigated in the future.    

\begin{figure}[!ht] 
 \centering
  \includegraphics[totalheight=7.0cm]{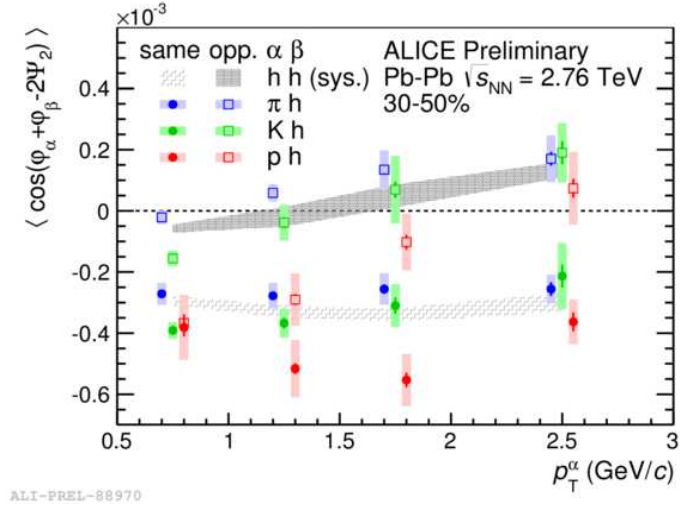}
  \caption{\small{Single identified opposite-sign (open) and same-sign
      (filled) $\gamma_{112}$ vs. average transverse momentum. The
      error bars and bands corresponds to statistical and systematic uncertainty
      respectively}}
 \label{fig:C112PID}
 \end{figure}

Fig.~\ref{fig:C112PID} shows the single identified $\gamma_{112}$  as
a function of the average transverse momentum for the 30--50\% central  
events in Pb--Pb  at  $\sqrt{s_{\rm NN}}$ = 2.76 TeV. 

The $\gamma_{112}$ with one identified pion which is found to be
similar to that for inclusive charge particles (shown by solid horizontal 
bands). The results for protons indicate a particle type dependence
although the large systematic uncertainty prevents a conclusion. With
the new data expected for Run 3 and Run 4 at the LHC, we can
significantly reduce the statistical/systematic uncertainty for
$\gamma_{112}$ correlations with one and two identified hardons.

\section{Summary}

We have presented the $\gamma_{112}$ and $\gamma_{132}$ correlator as a
function of centrality in Pb--Pb collisions at $\sqrt{s_{\rm NN}}$
= 5.02 TeV.  A comparison of $\gamma_{112}$  with earlier results
(Pb--Pb collisions at $\sqrt{s_{\rm NN}}$ = 2.76 TeV) shows no
significant increase in the signal or the background. The $\gamma_{132}$
correlator from data has been compared with a prediction from a Blast-wave 
+ LCC model (which is tuned to reproduce the measured elliptic flow in
data) as an attempt to estimate the background to
$\gamma_{112}$. Further improvement of the  Blast-wave 
+ LCC model is ongoing to describe other mixed and higher harmonic 
correlators ($e.g.$ $\gamma_{123}$ and $\gamma_{224}$) and estimate the 
CME contribution to  $\gamma_{112}$. A single identified $\gamma_{112}$
correlator is presented for mid central (30--50\%) events in Pb--Pb
collisions at $\sqrt{s_{\rm NN}}$ = 2.76 TeV. A slight hint of particle
type dependence is observed albeit with a large systematic
uncertainty. The larger data sample from ALICE in the Run 3 and
Run 4 period would allow us to draw more definitive conclusions about 
particle-type dependence of $\gamma_{112}$.

\end{document}